\documentclass{elsart5}

\usepackage{graphicx}
\usepackage{amsmath,amssymb}
\usepackage{threeparttable}

\usepackage[hang]{subfigure}
\usepackage[numbers,square,sort&compress]{natbib}
\usepackage[colorlinks=true,citecolor=blue,linkcolor=blue]{hyperref}%

\begin{document}

\begin{frontmatter}

\title{Phase stability of hafnium oxide and zirconium oxide on silicon substrate}

\author{Dongwon Shin\corauthref{cor1}}
\ead{dus136@psu.edu} \corauth[cor1]{Corresponding author.}
\author{Zi-Kui Liu}%
\address{Department of Materials Science and Engineering,\\
The Pennsylvania State University, University Park, PA 16802, USA}

\date{\today}%

\begin{abstract}
Phase stabilities of Hf-Si-O and Zr-Si-O have been studied with
first-principles and thermodynamic modeling. From the obtained thermodynamic
descriptions, phase diagrams pertinent to thin film processing were calculated.
We found that the relative stability of the metal silicates with respect to
their binary oxides plays a critical role in silicide formation. It was
observed that both the HfO$_2$/Si and ZrO$_2$/Si interfaces are stable in a
wide temperature range and silicide may form at low temperatures, partially at
the HfO$_2$/Si interface.
\end{abstract}


\end{frontmatter}

%
The thickness of SiO$_2$ as a gate oxide material in advanced complementary
metal oxide semiconductor (CMOS) integrated circuits has continuously decreased
and reached the current processing limits\cite{2001Wil}. Alternative materials
with higher dielectric constants, such as HfO$_2$ and ZrO$_2$, are considered
as candidates to replace SiO$_2$ for further improvement of their
performance\cite{1996Hub}. However, during the thin film deposition or the
subsequent rapid thermal annealing, oxides, silicates, and silicides may form
at the interface since most high-$k$ materials are metal
oxides\cite{2000Cop,2002Gut}. Among those interfacial phases, silicides are
detrimental to capacitor performance due to their metallic
behavior\cite{2005Cho}. In this regard, thermodynamic stability calculations
and experimental results have shown that the interface between HfO$_2$ and Si
is found to be stable with respect to the formation of silicides\cite{2002Gut}.
On the other hand, the ZrO$_2$/Si interface was found to be unstable around
1000K, which is in contradiction to the calculation by \citet{1996Hub}. It was
also observed that the Hf-silicide forms upon decomposition of HfO$_2$ in low
oxygen partial pressures \cite{2003Wan,2005Miy,2005Cho,2005Tak2} and HfSiO$_4$
suppresses Hf-silicide formation\cite{2005Tak}.

Although the phase stabilities in the Hf-Si-O and Zr-Si-O systems are
important, comprehensive thermodynamic explanations are not yet available. In
this paper, based on the recently developed thermodynamic descriptions of the
Hf-Si-O\cite{2006Shi} and Zr-Si-O systems with first-principles calculations
and thermodynamic CALculation of PHAse Diagrams
(CALPHAD)modeling\cite{1998Sau}, various phase diagrams pertinent to thin film
processing are investigated.

%
In the CALPHAD approach, the Gibbs energies of individual phases in a system
are evaluated from the existing experimental data with the so-called sublattice
model based on the crystal structures. The Gibbs energies of a higher-order
system can be readily extrapolated from the lower-order systems, and any new
phases of the higher-order system can be introduced. However, it is not always
possible to have enough experimental data for thermodynamic modeling of a
system\cite{1996Hub} so that theoretical calculations, such as first-principles
calculation results, can be used as supplementary experimental data. The
Hf-Si-O system was recently modeled with first-principles calculations and the
CALPHAD approach\cite{2006Shi}. The formation enthalpy for HfSiO$_4$ is
calculated from first-principles calculations since no experimental measurement
is reported. The reference states of the formation enthalpy for HfSiO$_4$ are
derived from the two binary metal oxides as shown in Eqn. \ref{eqn:hfsio4},
where $E$ represents the total energy of each phase. The formation entropy of
HfSiO$_4$ was evaluated from the temperature of peritectic reaction, HfO$_2$ +
Liquid $\rightarrow$ HfSiO$_4$, in the HfO$_2$-SiO$_2$ pseudo-binary system.
The thermodynamic description of the Zr-Si-O system was obtained by combining
the previous modelings\cite{1992Hal,1994Gue,2005Wan} and first-principles
calculation of ZrSiO$_4$ in the present work.

\begin{equation}\label{eqn:hfsio4}%
\Delta H^{\rm HfSiO_4}_f = E({\rm HfSiO_4}) -\tfrac{1}{2}E({\rm HfO_2})
-\tfrac{1}{2}E({\rm SiO_2})
\end{equation}

The highly efficient Vienna \emph{Ab initio} Simulation Package
(VASP)\cite{1996Kre} was used to perform the density functional theory (DFT)
electronic structure calculations. The projector augmented wave (PAW)
method\cite{1999Kre} was chosen, and the generalized gradient approximation
(GGA)\cite{1992Per} was used to take into account exchange and correlation
contributions. An energy cutoff was constantly set as 500 eV for all the
structures, and the Monkhorst-Pack scheme was used for the Brillouin-zone
integrations. For the $k$-point sampling, authors aimed all the structures to
have the $k$-point meshes as close as $ ({\rm \#~of~atoms~in~a~structure})
\times k_x\times k_y \times k_z \simeq 5000$ $k$-points. Thus, HfSiO$_4$ and
ZrSiO$_4$, for example, have $8\times 8\times 8$ $k$-point meshes. The
calculated results of metal oxides and silicates are listed in Table
\ref{tbl:abinitio}.

\begin{table*}[htb]
\centering%
\fontsize{8}{8pt}\selectfont %
\caption{First-principles calculation results of metal oxides and metal
silicates.}
\label{tbl:abinitio}%
\begin{tabular}{llccccccrcc}
\hline %
Phases & Space & \multicolumn{6}{c}{Lattice parameters}    & Total energy %
& $\Delta H_f$$^a$ & $\Delta S_f$$^{a,b}$\\
       & Group & a & b & c & $\alpha$ & $\beta$ & $\gamma$ & (eV/atom)    %
& kJ/mol-atom & J/mol-atom$\cdot$K\\
\hline %
HfO$_2$   & $P2_1/c$   & 5.135 & 5.194 & 5.314 & 90 & 99.56 & 90 & -10.2101 & - & -\\
ZrO$_2$   & $P2_1/c$   & 5.221 & 5.287 & 5.398 & 90 & 99.63 & 90 & -9.5376  & - & -\\
SiO$_2$   & $P3_221$   & 5.007 & 5.007 & 5.496 & 90 & 90 & 120 & -7.9581  & - & -\\
HfSiO$_4$ & $I4_1/amd$ & 6.616 & 6.616 & 6.004 & 90 & 90 & 90 & -9.1024  & -1.769 & -0.219\\
ZrSiO$_4$ & $I4_1/amd$ & 6.698 & 6.698 & 6.038 & 90 & 90 & 90 & -8.7723  & -2.358 & -0.275\\
\hline %
\end{tabular}
\begin{tablenotes}
\item $\rm ^a$ Formation enthalpies and entropies of metal silicates are
expressed with respect to their binary oxides.
\item $\rm ^b$ Formation entropies are evaluated from temperature of peritectic
    reactions ($M$O$_2$ + Liquid $\rightarrow$ $M$SiO$_4$).
\end{tablenotes}
\end{table*}

\begin{figure}[htb]
\centering %
    \subfigure[Isopleth of HfO$_2$ and Si] {
        \label{fig:isophf}%
        \includegraphics[width=3.35in]{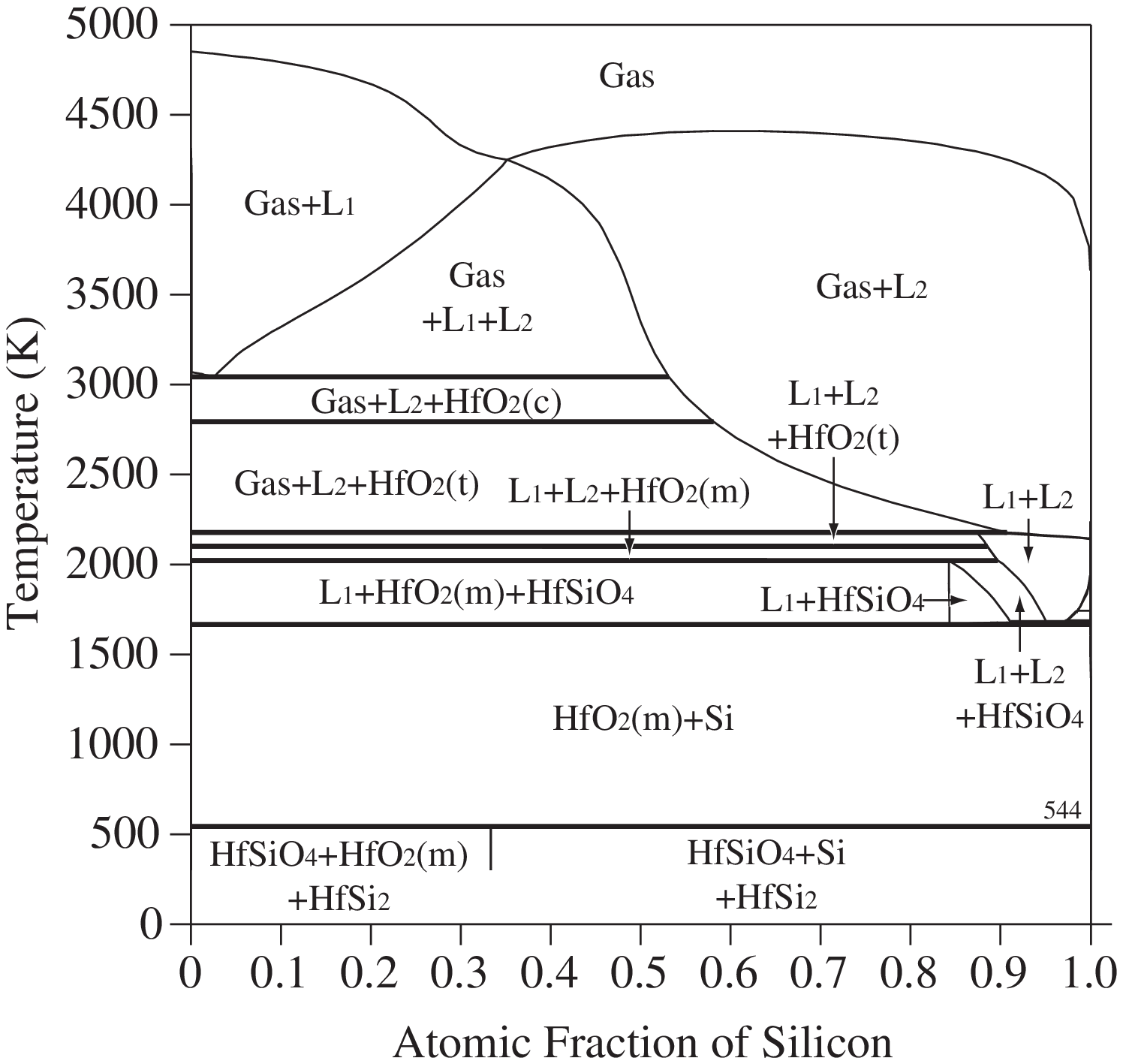}
    }
    \subfigure[Isopleth of ZrO$_2$ and Si] {
        \label{fig:isopzr}%
        \includegraphics[width=3.35in]{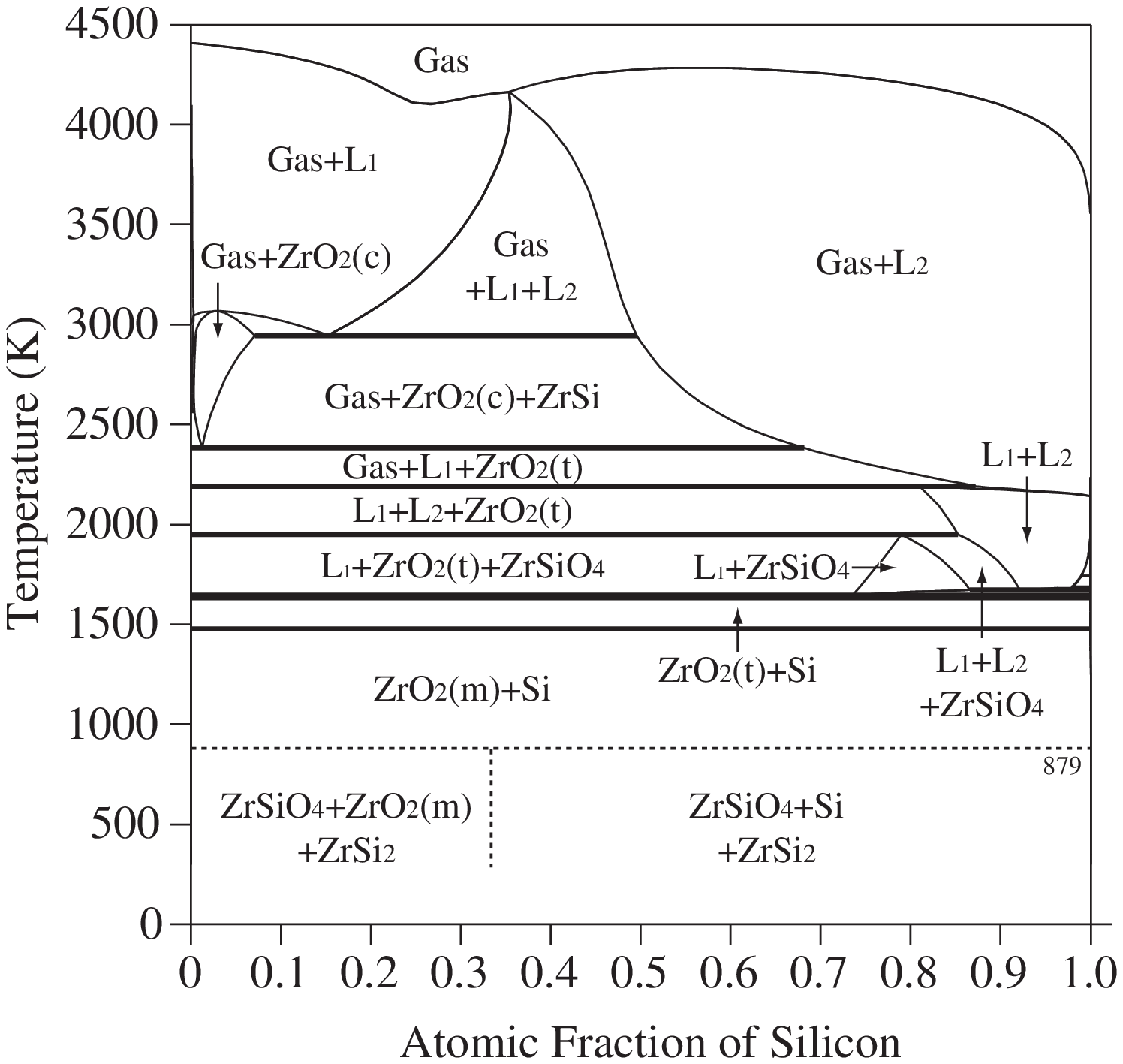}
    }
\caption{%
Calculated isopleths of HfO$_2$-Si and ZrO$_2$-Si at 1 atm. Polymorphs of metal
oxides for HfO$_2$ and ZrO$_2$, i.e. monoclinic, tetragonal, and cubic, are
given in parentheses.
}%
\label{fig:isopleth} %
\end{figure}

%
From the constructed thermodynamic databases of the Hf-Si-O and Zr-Si-O
systems, the isopleths of HfO$_2$-Si and ZrO$_2$-Si are calculated in order to
investigate the stability range of silicides at the metal oxides/silicon
interface and are given in Figure \ref{fig:isopleth}. Calculated results show
that HfSi$_2$ is stable up to 544K based on the formation enthalpy of HfSiO$_4$
from first-principles calculations. It is generally accepted that the
uncertainty of the formation enthalpy of intermetallic compounds, which
originates from the density functional theory itself, is about $\pm$1
kJ/mol-atom\cite{2002Wol,1998Fri}. Thus the associated decomposition
temperature of HfSi$_2$ at the HfO$_2$/Si interface ranges from 382 to 670K
when the formation enthalpy of HfSiO$_4$ is adjusted within its uncertainty
range from --0.769 to --2.769 kJ/mol-atom. The formation entropy of HfSiO$_4$
with respect to the binary oxides was evaluated correspondingly to reproduce
its peritectic reaction at 2023K. It should be noted that the phase stability
range of HfSi$_2$ in the HfO$_2$-Si isopleth is not directly correlated with
the first-principles calculation of HfSiO$_4$, but predicted from the Gibbs
energies of other phases, including the HfSiO$_4$ phase, in the Hf-Si-O system.
Even with the uncertainty of formation enthalpy for HfSiO$_4$, the temperature
range for the HfO$_2$ and Si coexistent phase region in the isopleth is fairly
wide from 670 to 1700K.

For ZrSiO$_4$, besides the uncertainty of formation enthalpy from
first-principles in the present work, the peritectic reaction (ZrO$_2$ + Liquid
$\rightarrow$ ZrSiO$_4$) temperature in the ZrO$_2$-SiO$_2$ pseudo-binary is
also uncertain from 1910 to 1949K. Thus, the formation entropy of ZrSiO$_4$
varies accordingly. The Gibbs energy of ZrSiO$_4$ at 1000K evaluated from the
formation enthalpy derived from first-principles and formation entropy
evaluated from the peritectic temperature of 1949K (listed in Table
\ref{tbl:abinitio}) is almost identical to the value used by \citet{1996Hub}.
With these formation enthalpy and entropy values of ZrSiO$_4$, ZrSi$_2$ is
completely suppressed by ZrSiO$_4$ and does not show up in the ZrO$_2$-Si
isopleth. To make ZrSi$_2$ appear in the ZrO$_2$-Si isopleth, the formation
enthalpy of ZrSiO$_4$ should be more negative than the first-principles
calculation result within the uncertainty of formation enthalpy and peritectic
temperature for ZrSiO$_4$. When formation enthalpy of ZrSiO$_4$ with respect to
the binary metal oxides is set to its lowest limit from the uncertainty of
first-principles calculations, $\Delta H^{\rm ZrSiO_4}_f=-3.358$ kJ/mol-atom,
and entropy of formation is evaluated as $\Delta S^{\rm ZrSiO_4}_f=0.788$
J/mol-atom$\cdot$K, ZrSi$_2$ is stable up to 879K in the ZrO$_2$-Si isopleth.
Then formation enthalpy of ZrSiO$_4$ is $\Delta H^{\rm ZrSiO_4}_f=-338.568$
kJ/mol-atom with respect to SER (Standard Element Reference) and this agrees
well with the experimental measurement, --339.033 kJ/mol-atom from
\citet{1992Ell}. Consequently, the \emph{safe} temperature range for ZrO$_2$ to
be stable with Si is between 879 and 1630K, narrower than that of HfO$_2$ and
Si. However, even with these uncertainties, both metal oxides are stable with
Si approximately above 900K as summarized by \citet{1996Hub}(1000K).

These isopleth calculation results are in agreements with \citet{2002Gut} for
the Hf-Si-O system but not with the Zr-Si-O system. In their calculations, they
assumed that thermal effects are of secondary importance for Gibbs energy
change so that the contribution from entropy was ignored in the calculations
for silicide formation reactions. However, our calculations from the individual
thermodynamic databases showed that such an entropy effect cannot be neglected.
According to our calculation results, both HfO$_2$ and ZrO$_2$ are stable with
Si and this is in agreement with the calculation from \citet{1996Hub}. However,
ZrSi$_2$ was found at the metal oxide/Si interface in their
experiment\cite{2002Gut} while HfO$_2$ was stable on Si without any silicides
formation when they deposited at 823K and then annealed at 1073K. It can be
explained that since their fabrication process was rapid thermal chemical-vapor
deposition (RTCVD), it might not have reached the thermodynamic equilibrium
state. Furthermore, the oxygen partial pressure of their experiment was not
reported. The effect of oxygen partial pressure will be discussed later in this
paper.

The calculated isopleths indicate that metal silicates play an important role
in the silicide formation as suggested by \citet{2005Tak} From first-principles
calculations, the formation enthalpy of ZrSiO$_4$ is $-2.358\pm1$ kJ/mol-atom
whereas that of HfSiO$_4$ is only $-1.769\pm1$ kJ/mol-atom when the reference
states are set to the binary oxides. It is intriguing to see that such a small
(0.6 kJ/mol-atom) difference in the formation of metal silicates greatly
affects the phase stability at the metal oxides/silicon interface. This can be
explained by comparing the relationship between metal oxides, silicates, and
silicides in the isothermal section.

\begin{figure}[htb]
\centering %
    \subfigure[Zr-Si-O at 500K] {
        \label{fig:isot500}%
        \includegraphics[width=3.25in]{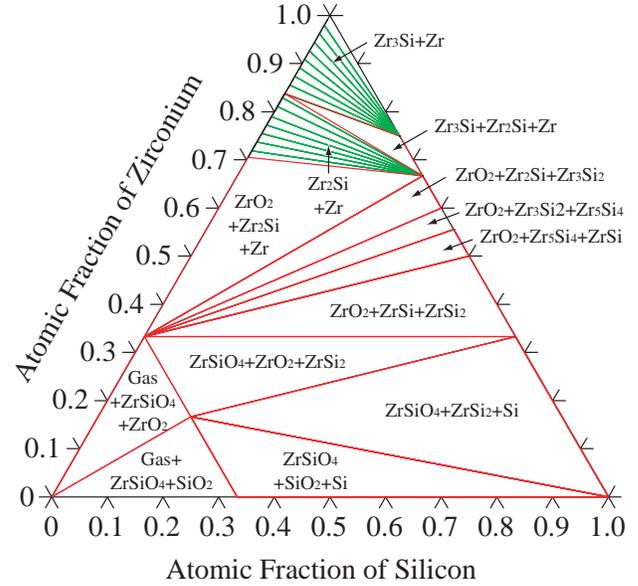}
    }
    \subfigure[Zr-Si-O at 1000K] {
        \label{fig:isot1000}%
        \includegraphics[width=3.25in]{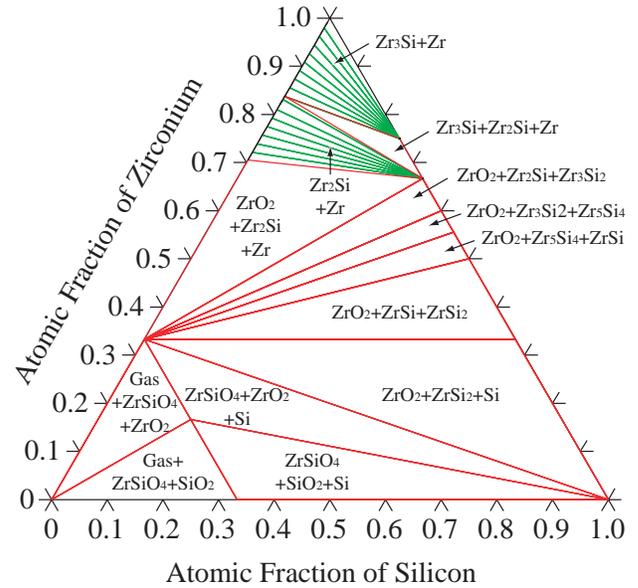}
    }
\caption{%
Calculated isothermal sections of Zr-Si-O at (a) 500K and (b) 1000K.
Tie lines are drawn inside the two phase regions.
}%
\label{fig:isot} %
\end{figure}

Isothermal sections of the Zr-Si-O system at two different temperatures, 500K
and 1000K, are calculated (see Figure \ref{fig:isot}) to investigate the phase
relationship regarding the decomposition of ZrSi$_2$ at the ZrO$_2$/Si
interface. The two different three-phase regions, ZrSiO$_4$+ZrO$_2$+ZrSi$_2$
and ZrSiO$_4$+ZrSi$_2$+Si, in the 500K isothermal section are intersected by
the line connecting ZrO$_2$ and Si. Therefore, ZrSi$_2$ can be found in the
thin film process. However, the 1000K calculation shows that ZrO$_2$ is stable
with the Si substrate without any silicide formation as there is a tie line
connecting ZrO$_2$ and Si. Isothermal sections of Hf-Si-O at the same
temperatures, 500K and 1000K, showed similar phase stabilities as
Zr-Si-O\cite{2006Shi}.

According to the calculation results of isopleths and isothermal sections of
the Hf-Si-O system, HfO$_2$ and Si should be stable at the temperature range
between 670K and 1700K. However, it is reported that under oxygen-deficient
conditions, Hf-silicide forms at the HfO$_2$/Si interface even in this
temperature range. \citet{2003Wan} found that oxygen-deficient HfO$_{x<2}$
consumes the oxygen in the SiO$_2$ thin layer covered on the silicon substrate
to form fully oxidized metal oxide. Even the layer of silicates will be
decomposed along with the progress of HfO$_{x<2}$ deposition. The recent work
from \citet{2005Miy} also confirmed the formation of nanometer-scale HfSi$_2$
dots on the newly opened void surface produced by the decomposition of
HfO$_2$/SiO$_2$ films at the oxide/void boundary in vacuum.

\begin{figure}[htb]
\centering %
    \subfigure[$x_{\mathrm{Hf}}/x_{\mathrm{Si}}$=1] {
        \label{fig:hfpo2}%
        \includegraphics[width=3.25in]{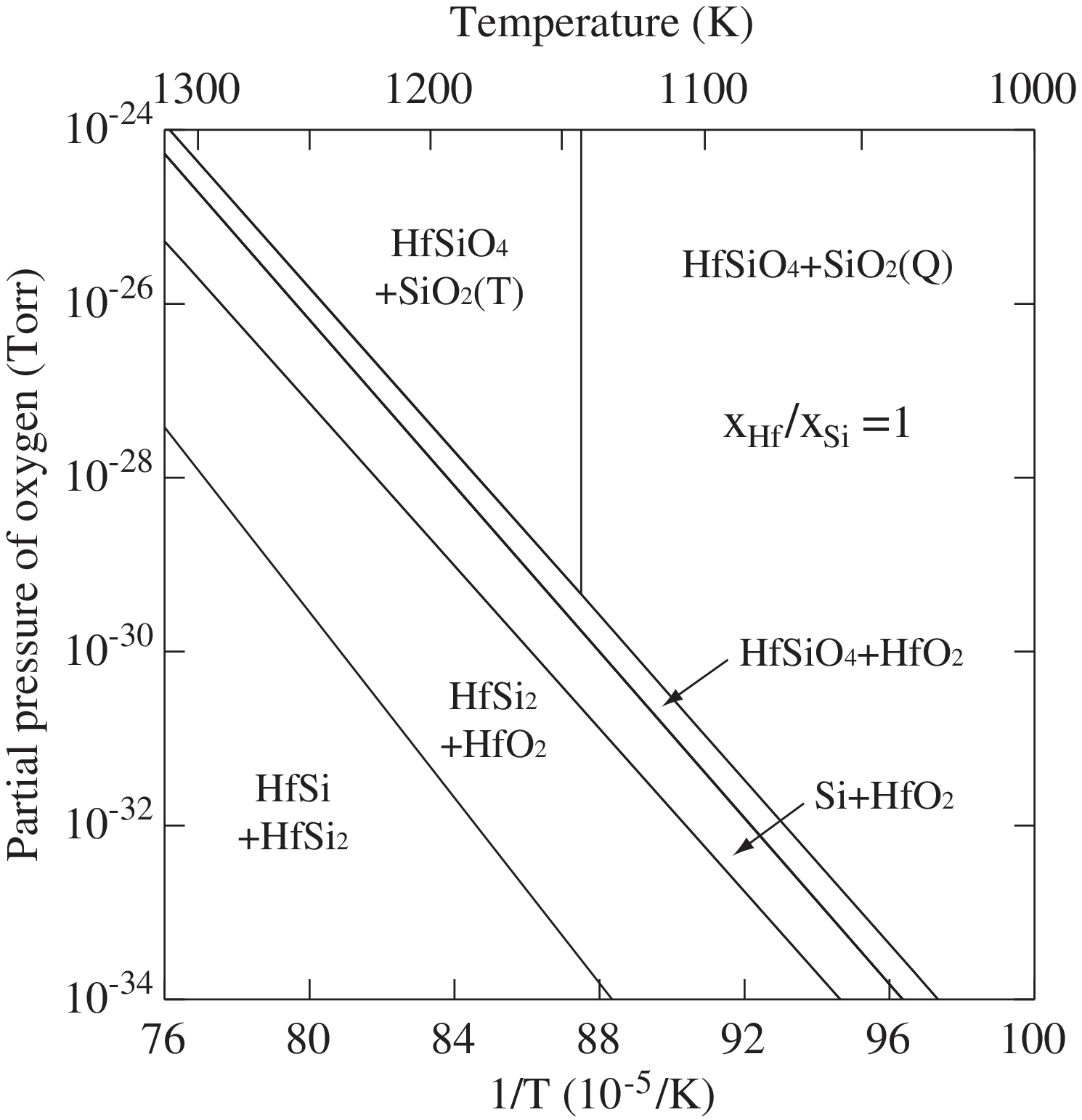}
    }
    \subfigure[$x_{\mathrm{Zr}}/x_{\mathrm{Si}}$=1] {
        \label{fig:zrpo2}%
        \includegraphics[width=3.25in]{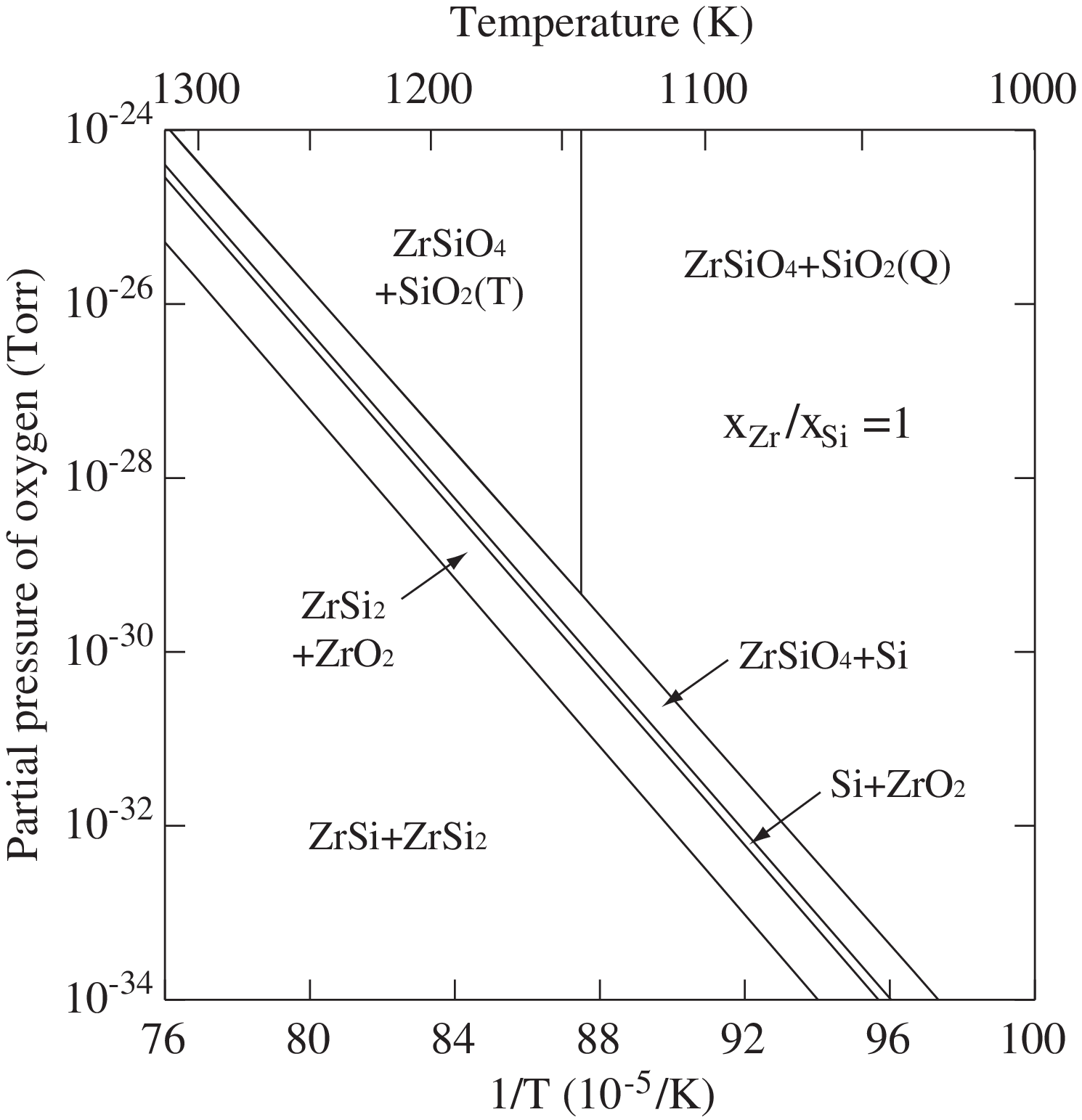}
    }
\caption{%
Partial pressure of oxygen vs. temperature phase diagrams for (a) Hf-Si-O and
(b) Zr-Si-O systems when $x_{\rm Hf,Zr}$/$x_{\rm Si}$=1. SiO$_2$(Q) and
SiO$_2$(T) represent Quartz and Tridymite, respectively.
}%
\label{fig:po2} %
\end{figure}

To further understand the effect of the oxygen partial pressure, the phase
diagrams of oxygen partial pressure-temperature are calculated and shown in
Figure \ref{fig:po2} with the ratio between the metals (Hf and Zr) and Si set
to 1. It should be mentioned here that the oxygen partial pressure in these
calculations are the \emph{local} oxygen pressure at the interface, which is
extremely low. Consequently, both systems initially form only metallic
silicides. As oxygen partial pressure increases, part of the silicides
transform into metal oxides. Afterwards, HfO$_2$ and Si are stable in the
Hf-Si-O system as confirmed by
experiments.\cite{2003Wan,2005Miy,2005Cho,2005Tak2} Then, HfO$_2$ is in
equilibrium with HfSiO$_4$. In the Zr-Si-O system, with further oxidization,
ZrO$_2$ is stable with Si. However, the phase region is narrower than that of
the Hf-Si-O system. This is in agreement with \citet{2000Cop} that ZrO$_2$ is
vulnerable to high temperature vacuum annealing. Therefore, it is possible to
have a stable ZrO$_2$/Si interface, but this is very challenging in the high
vacuum condition.

In summary, with the thermodynamic descriptions of the Hf-Si-O and Zr-Si-O
systems developed by the CALPHAD technique, isopleths and isothermal sections
can be readily calculated. It is found that the HfO$_2$/Si interface is
thermodynamically stable between 670 and 1700K as far as oxygen partial
pressure is high enough to keep HfO$_2$ stable. ZrO$_2$/Si interface is stable
between 879 and 1630K, but in the oxygen-deficient condition, the processing
window for a stable ZrO$_2$/Si interface is very narrow. Both metal oxides are
stable with a Si substrate above 900K, even with the uncertainties of the
formation enthalpies and entropies for metal silicates.

This work is funded by the National Science Foundation (NSF)
through grants DMR-0205232 and 0510180. First-principles calculations were
carried out on the LION clusters at the Pennsylvania State University supported
in part by the NSF grants (DMR-9983532, DMR-0122638, and DMR-0205232) and in
part by the Materials Simulation Center and the Graduate Education and Research
Services at the Pennsylvania State University.

\bibliographystyle{unsrtnat}

\end{document}